# An asymmetric surface coating strategy for promotes rapid endothelialization in the rabbit carotid artery

Lili Tan[1,2], Zhiyi Ye[1], Suhua Yu[1], Jinxuan Wang[1,3], Chenxi Ouyang[4], Zhengcai Zhang[5], Robert Guidoin[6], Guixue Wang[1,2*]

[1] Key Laboratory for Biorheological Science and Technology of Ministry of Education, National Local Joint Engineering Lab for Vascular Implants, College of Bioengineering, Chongqing University, Chongqing 400044, China.

[2] JinFeng Laboratory, Chongqing 401329, China

[3] School of Basic Medicine, Chengdu Medical College, Chengdu 610500, China

[4] Department of Vascular Surgery, Fuwai Hospital, National Center for Cardiovascular Diseases, Chinese Academy of Medical Sciences and Peking Union Medical College, Beijing 100037, China.

[5] Lepu Medical Technology (Beijing) Co.,Ltd., Beijing 102200, China

[6] Department of Surgery, Laval University, Quebec, Canada

*Corresponding author: wanggx@cqu.edu.cn (GX Wang)

**Abstract:** Studying surface modification has long been a key area for enhancing the effects of vascular stents after surgery. The study aimed to develop an asymmetric drug-eluting stent (ADES) with differential drug loading on its inner and outer surfaces, hypothesizing that this design would enhance drug delivery efficacy for percutaneous coronary interventions (PCIs) compared to uniformly coated drug-eluting stents (UDES). An ultrasonic atomization spraying device was utilized to fabricate the ADES, which was subsequently evaluated for drug release patterns, hemocompatibility, and biocompatibility. In vitro, assessments demonstrated favorable hemocompatibility and showed targeted drug delivery capabilities of ADES within artificial blood vessels. Furthermore, in vivo testing using a rabbit carotid artery model revealed significant endothelialization on stented segments treated with the ADES. These findings suggest that the ADES holds promise as a minimally invasive platform for improving cardiovascular disease treatment outcomes by addressing thrombus formation and neointima proliferation more effectively than traditional stents.

**Keywords:** asymmetric surface, ultrasonic atomizing spraying, endothelialization,drug eluting stent.

# 1 Introduction

Coronary artery disease (CAD) remains a leading cause of morbidity and mortality worldwide, necessitating the development of advanced therapeutic interventions.[1] Percutaneous coronary intervention (PCI), especially with drug-eluting stents (DES), has revolutionized the treatment landscape by reducing restenosis rates compared to bare-metal stents (BMS).[2] However, despite these advancements, DES still faces challenges such as incomplete endothelialization,[3, 4] late stent thrombosis,[5, 6] and the need for prolonged dual antiplatelet therapy,[7] all of which can compromise patient outcomes.

Traditional DES designs apply a uniform coating of antiproliferative drugs on both the abluminal (outer) and luminal (inner) surfaces of the stent. While this approach effectively inhibits neointimal hyperplasia, it does not fully address the distinct biological environments encountered by the two surfaces post-implantation. The luminal surface is in direct contact with circulating blood, where rapid endothelialization is crucial to prevent thrombus formation and promote vascular healing. Conversely, the abluminal surface interacts with the vessel wall, where excessive extracellular matrix deposition and chronic inflammation can lead to adverse remodeling and stent malapposition.[8, 9]

Given these differing requirements, there is a growing interest in developing an asymmetric surface coating strategy that tailors drug delivery to each stent surface, thereby optimizing therapeutic efficacy while minimizing side effects.[10] By loading distinct drugs onto the inner and outer surfaces, the asymmetric surface coating strategy aims to enhance endothelialization on the luminal side while controlling smooth muscle cell proliferation on the abluminal side without compromising hemocompatibility or biocompatibility.

This project aims to create an asymmetric drug-eluting stent (ADES) using a novel ultrasonic atomizing spraying technology for fine control over drug distribution and release kinetics. We hypothesize that this personalized drug coating method will outperform standard uniform-coating DES (UDES) by inhibiting neointimal development more effectively and promoting faster, more complete endothelial coverage. Through in vitro studies of hemocompatibility, in-depth investigation of drug release patterns, and thorough evaluation in animal models, we hope to verify the potential of ADES as a next-generation platform for cardiovascular therapies, giving improved outcomes with less invasiveness.

## 2 Materials and methods

### 2.1 Stent preparation

An ultrasonic atomization spraying device developed in-house by the laboratory[11, 12] was used to prepare the ADES and UDES samples for this study. During the spraying process, the 316 L stainless steel BMS was cleaned by an ultrasonic cleaning instrument. Fibronectin (FN) and paclitaxel (PTX) were used to coat the inner wall and outer wall of the BMS, respectively. The spraying concentration of FN was 10 μg/ml, and the PTX working concentration was 1 mg/ml. In detail, an appropriate amount of FN or PTX was weighed and dissolved in a 1 mg/ml solution of PLLA. The operating conditions of the spraying device were as follows: carrier air pressure of 1 psi, spraying time of 40 s, and rotational advancement of the BMS at a speed of 2 r/s. To confirm the successful fabrication of ADES or UDES, the chitosan/FN solution was combined with rhodamine B (RB) and sprayed across the whole surface of the stent. After sealing the stent's inner wall with a balloon, the exterior surface was sprayed with a PLLA/PTX solution labeled with fluorescein isothiocyanate (FITC).

**2.2 Hemolysis assay**

The ADES, UDES and BMS samples were added separately to a test tube containing 10 ml of saline and incubated at 37°C for 30 min. Blood from healthy volunteers was taken from the hospital and diluted with saline at a ratio of 4:5, and the pretreated samples were removed, placed into sterile 1.5 ml EP tubes, labeled, added to the diluted blood, and incubated for 60 min at 37°C in an incubator. The negative control was saline, and the positive control was distilled water. Next, the tubes were centrifuged for 5 min (centrifugation conditions: 3000 rpm/min), distilled water was used as a control, and the absorbance of the collected supernatant was measured at a wavelength of 545 nm. The hemolysis rate was calculated as follows (n=3): % hemolysis = [(OD sample - OD negative)/(OD positive - OD negative)] × 100%.

**2.3 APTT, PT and TT tests**

Activated partial thromboplastin time (APTT), a commonly used indicator of the endogenous coagulation system, is often used to detect endogenous coagulation; prothrombin time (PT) and thrombin time (TT) are commonly used indicators of the exogenous coagulation system and are often used to detect exogenous coagulation. The detailed descriptions in these papers are needed to guide specific experimental steps.[13]

**2.4 Cell seeding and roller culture**

VSMCs were recovered from the laboratory cell repository. VSMCs were cultured in DMEM (BI) supplemented with 10% fetal bovine serum (GIBCO) under standard cell culture conditions (37°C, humidified, 5% CO2/95% air environment). The medium was replenished every other day during cell culture. Cells in the logarithmic growth phase were subjected to pancreatic enzyme digestion and single-cell suspension, at a concentration of 2×104/ml. Then, the same amount of cells was added to the rotating culture tube with our support. Before culturing, the rotary cell culture system was sterilized by ultraviolet irradiation for half an hour.

**2.5 Immunofluorescence staining**

Stent samples were collected after 3 days of rolling culture and fixed with 4% paraformaldehyde for 15 min, followed by permeabilization in PBS supplemented with 0.25% Triton X-100 for 10 min at room temperature. The cells were then incubated with FITC phalloidin for 1 hour at 37°C, washed three times with PBS, and mounted with DAPI (Thermo Fisher, USA). The stents were photographed using a fluorescence microscope. Images of at least three independent random fields per sample were acquired from three independent experiments.

**2.6 Evaluation of drug desolvation from the stents**

The stents were integrated into a vessel-like chamber, which contained an agarose-cast tube mimicking the vascular wall. This chamber was perfused with a phosphate buffer solution (PBS) reservoir. The flow rate and pulse cycle were controlled by a peristaltic pump. All components were connected in series using sterile silicone tubing. This configuration differs from traditional stent drug release detection devices, allowing for separate detection of drug release from both the inner and outer walls of the stent. Sampling time points were set at 3 hours, 6 hours, 12 hours, 1 day, 2 days, 3 days, 5 days, 7 days, 14 days, and 28 days, with n=3 for each time point. The flow velocity was maintained at 15 dyn/cm$^2$. The agarose-casted tubes were collected, pulverized, and centrifuged at 10,000 rpm. The supernatant was used for drug content testing. Luminal fluid was also collected. An ELISA kit (BJ-E0494) was used to test for FN release following the kit instructions. The PTX release rate was detected using HPLC (Oxford, UK, LC600B).

### 2.7 Stent implantation and sample collection

Fifty New Zealand white rabbits (male, approximately 2.5 kg) were selected and kept in the Experimental Animal Center of Chongqing Xinqiao Hospital, where they were housed in a standard environment for one week of observation. The experimental animals were found to be healthy enough for use in subsequent experiments. Four days before the operation, aspirin tablets and clopidogrel were added to the feed at 2 mg/kg and 1.5 mg/kg, respectively. The experimental groups consisted of the ADES group and UDES group, while the control group was composed of the 316 L stainless steel bare stent group. The methods used for stent implantation and sample collection were described in papers published by the laboratory.[11, 14] The time points were set as 1/2/4/12 weeks, with three parallel samples collected at each cutoff point. At all time points, all experimental animals remained alive. The stent-implanted segment vessels were removed, fixed with 2.5% glutaraldehyde, longitudinally sliced, and then freeze-dried for scanning electron microscopy (SEM, VegaIII-LMH) observation and energy dispersive spectrometry (EDS, INCA x-sight 7557). Additionally, the main internal organs (heart, liver, spleen, lungs, and kidneys) were fixed with 4% paraformaldehyde, paraffin sectioned, and stained with hematoxylin and eosin (HE) for histological analysis.

### 2.8 Statistical analysis

The significance of variability among the means of the experimental groups was determined by 1-way or 2-way analysis of variance using SPSS 22.0. Differences among the experimental groups were considered to be statistically significant when at $p<0.05$. Unless indicated, values are given as the mean ± SD.

# 3 Results

## 3.1 Preparation of ADES and UDES

The preparation process of ADES and UDES is shown in the schematic diagram. The 316 L BMS was placed on the console of module 3. When the balloon works, the inner surface of the stent is segregated from the atomized aerosol, finally forming an AEDS carrying different drugs on the inner and outer walls. If the balloon does not work, a UDES is formed. In this spraying process, the outer wall of the stent was coated with PTX, an antiproliferation drug, while the inner wall was coated with FN, a protein that promotes reendothelialization; hypothetically, antiproliferative drugs target the vascular intima, whereas drugs that promote reendothelialization target the vascular endothelium (Figure 1B). Under a fluorescence microscope, red fluorescence was observed on the inner and outer walls of the ADES, while green fluorescence was observed only on the outer wall. Diverse red and green fluorescence was observed on both the inner and outer walls of the UDES. The observations suggested that only the outer wall of the ADES was coated with the PTX film, but both the inner and outer walls of the UDES were coated with PTX. Since then, we achieved the initial design concept and successfully constructed ADES and UDES for subsequent in vitro and in vivo experiments.

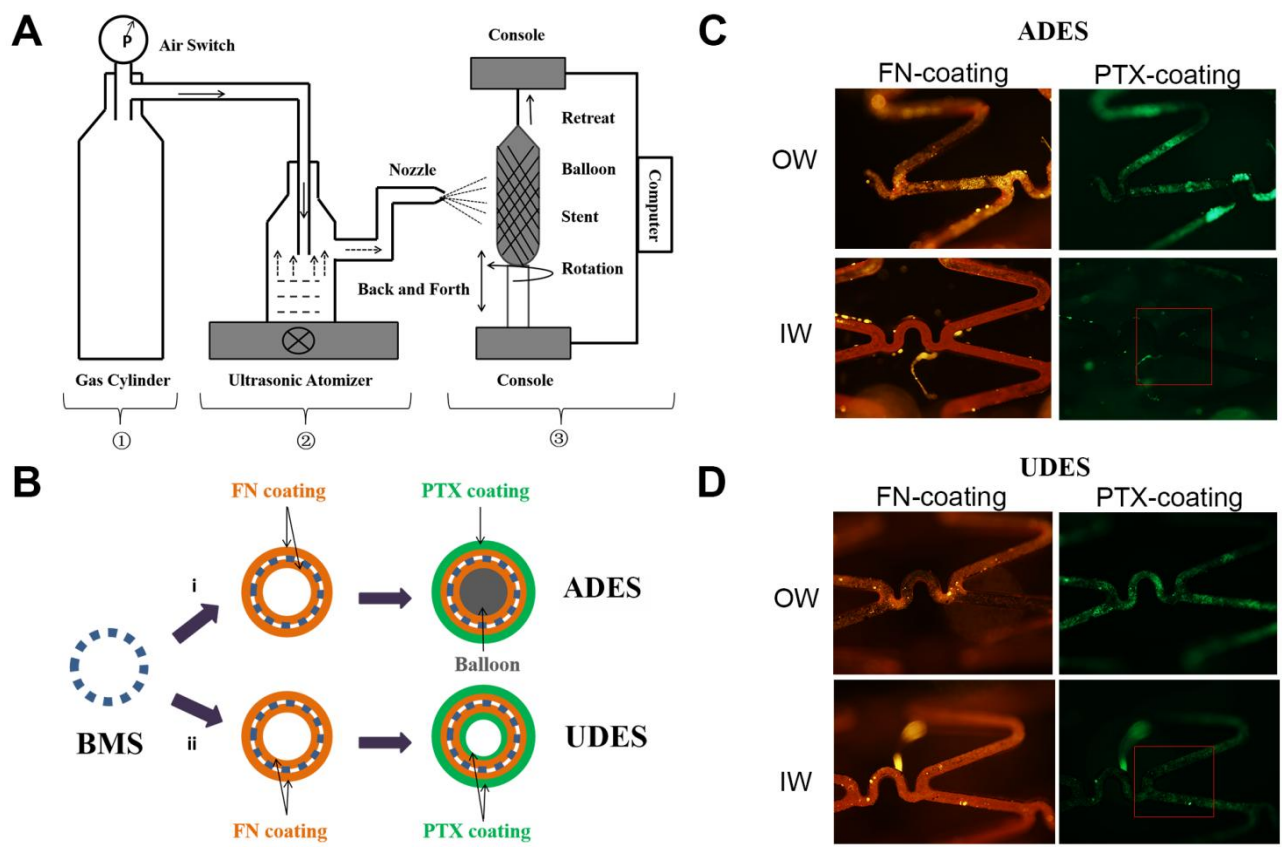

Figure 1. Design and preparation of ADES and UDES. (A) Schematic of the optimized ultrasonic spray stent system. ① The gas carrier module includes a nitrogen storage cylinder and an air switch. ② In the ultrasonic atomization module, the ultrasonic device atomizes the polymer solution. ③ The stent-carrier module, controlled by a computer, is composed of two consoles; one console connects the stent, and the other connects the balloon, stent and balloon. (B) Schematic design of ADES and UDES. (C, D) Fluorescence images of the inner and outer walls of the ADES and UDES, ×4 (Olympu, Japan), OW: outer wall, IW: inner wall.

### 3.2 The morphology of ADES and UDES

Compared with those of the BMS, the outer/inner side coatings of the ADES and UDES were globally smooth and uniform, and the enlarged image shows traces of coating on the surface of the stent. (Figure 2A). EDS showed that BMS contains Si, K, Cr, Mn, Fe, Ni, and other chemical elements and does not contain C or O (Fig. 2B), which indicates that the surface of BMS is not covered by organic matter coatings. In contrast, C and O were detected on both the inner and outer walls of the ADES and UDES. The weight percentages of C and O in the outer wall of the ADES are 17.45% and 9.23%, respectively, and those in the inner wall are 9.41% and 4.37%, respectively. In addition, the percentages of C and O in the outer wall of the UDES were 10.64% and 17.32%, respectively, and those in the inner wall were 8.62% and 18.25%, respectively. Obviously, the weight percentages of C and O on the inner wall of the ADES were lower than those on the UDES. These results further indicate that an asymmetric coating and a uniform coating were sprayed onto the surface of the bare metal scaffolds and that the amounts of elemental C and O on the outer walls of the ADES and UDES were not significantly different.

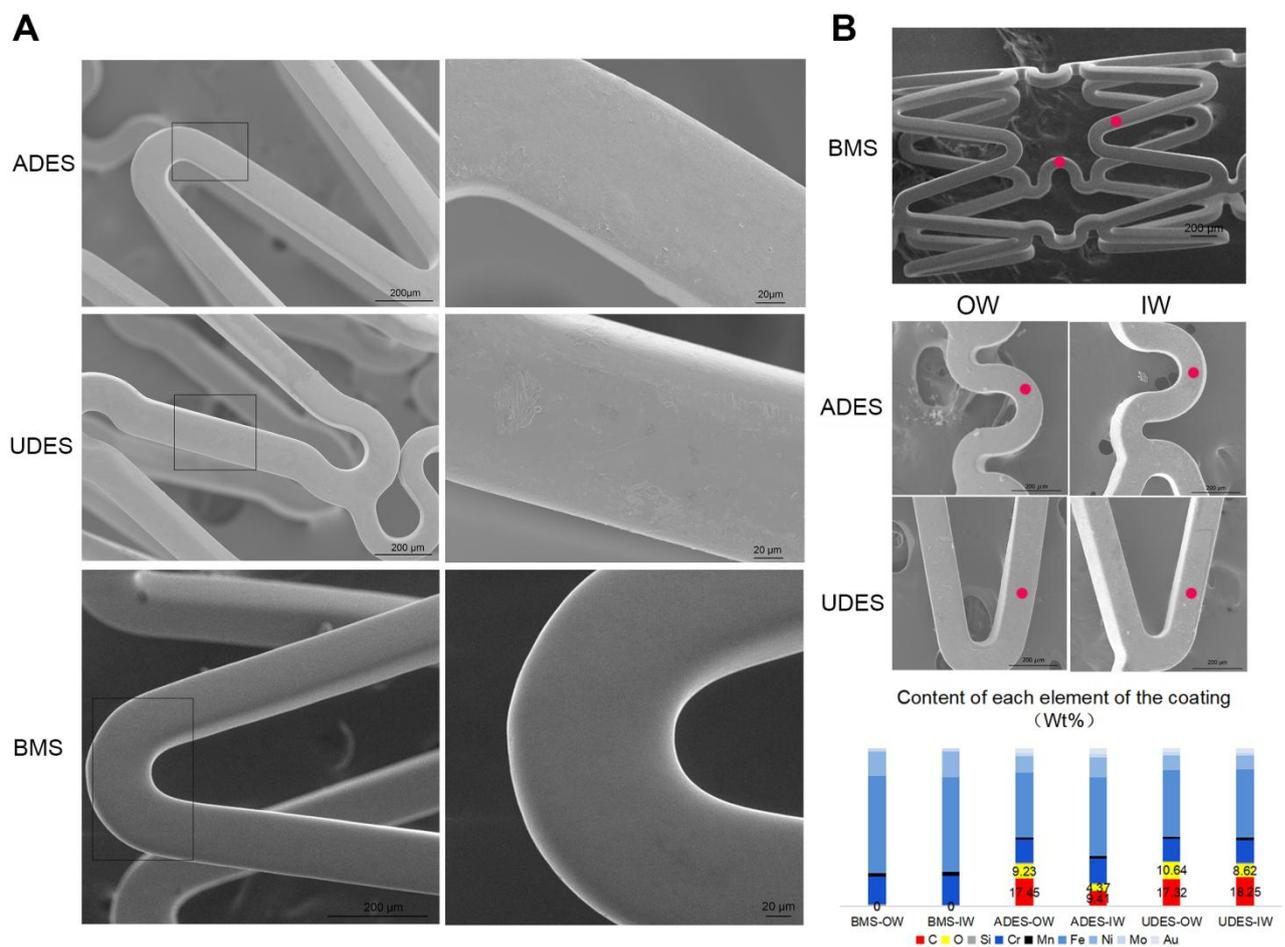

Figure 2. Surface morphology and energy spectroscopy of the stent coatings. (A) SEM images of ADES, UDES and BMS. (B) Coating surface of the EDS test. OW: outer wall, IW: inner wall.

### 3.3 Hematocompatibility and biosafety

As shown in Table 1, the hemolysis rates of BMS, ADES, and UDES were 1.78%, 2.82%, and

2.81%, respectively, which were all less than 5%, satisfying the biosafety evaluation reference standard for the hemolysis rate of medical implant materials.

The APTTs for BMS, ADES, and UDES were 27.61±0.4419 s, 26.40±0.7532 s, and 25.89±0.9434 s, respectively. The PTs were 12.02±0.3302 s, 11.37±0.6185 s, and 12.05±0.1693 s, respectively. TTs were 12.89±0.2216 s, 12.03±0.3705 s, and 11.97±0.1817 s, respectively. The data were not significantly different. (Figure 3A)

Pathologic images of the major organs after stent implantation are shown in Figure 3B. The pathological and histological morphology of the major organs did not change significantly, suggesting that there was no adverse effect on the major organs of the experimental animals after stent implantation. From the results of blood biochemistry tests, it can be concluded that the blood indices of experimental animals before and after stent implantation did not significantly change, suggesting that the blood compatibility and biosafety of ADES and UDES are good and that there is no statistically significant difference.

Table 1 The results of hemolysis rate testing

| Groups | Absorbance value (A) | Hemolysis ratio | Statistical differences |
|---|---|---|---|
| Positive control | 0.9847±0.0004 | | — |
| Negative control | 0.0077±0.0002 | | |
| BMS | 0.0251±0.0003 | 1.78% | |
| ADES | 0.0353±0.0004 | 2.82% | P>0.05 |
| UDES | 0.0351±0.0005 | 2.81% | |

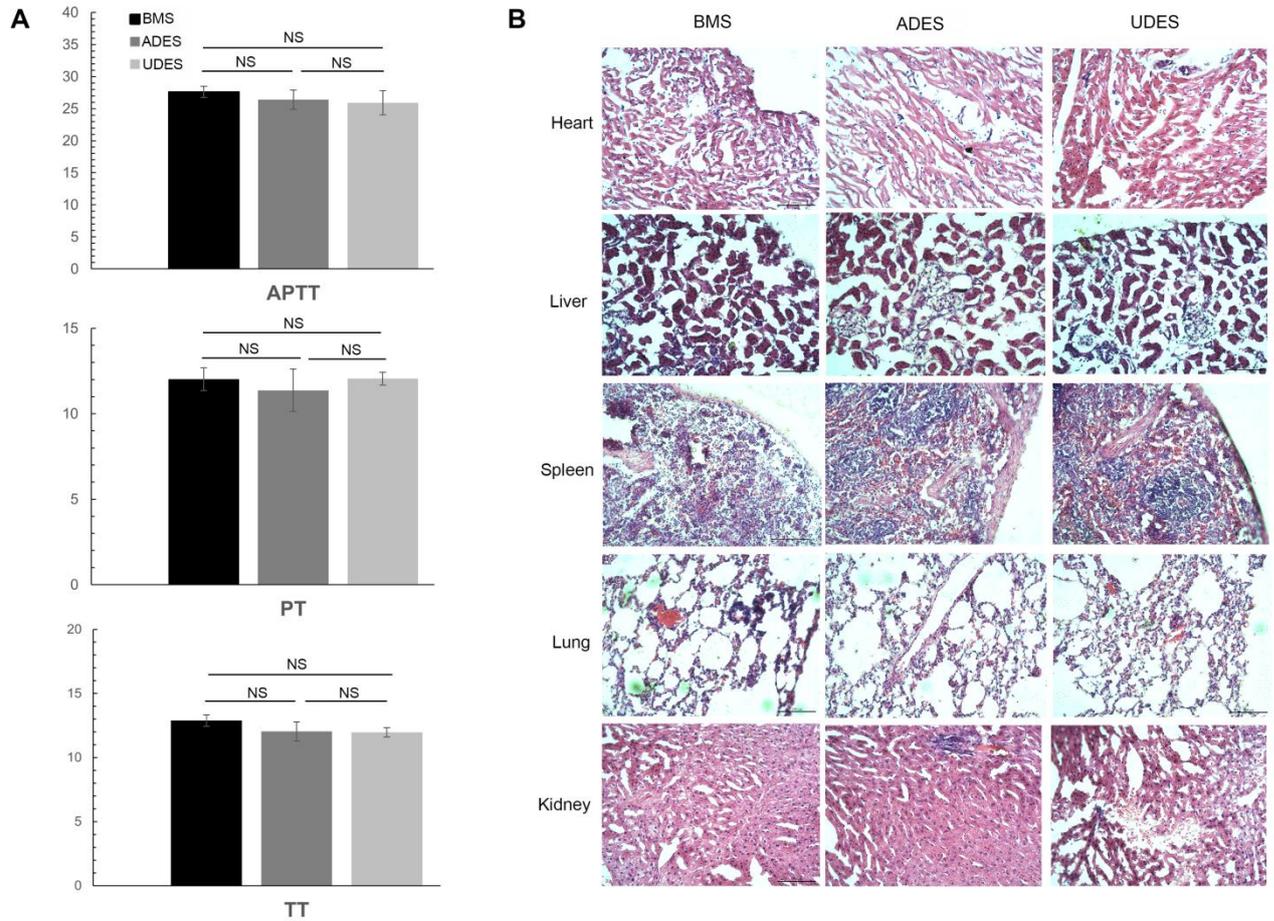

Figure 3. (A) APTT, PT, and TT for BMS, ADES, and UDES. The y-axis is the time in seconds. NS: no significance. (B) Pathologic images of the major organs after stent implantation.

**3.4 HUVEC migration on the FN and PTX coatings**

Compared with the bare slide group, the FN coating favored cell migration, and the PTX coating inhibited cell migration, as shown in Figure 4A. After statistically analyzing the migration area, as shown in Figure 4B, there were highly significant differences in the data between the groups at the three time-points of 4, 10, and 24 hours. The results demonstrated that FN promoted endothelial cell repair, while PTX inhibited cell repair.

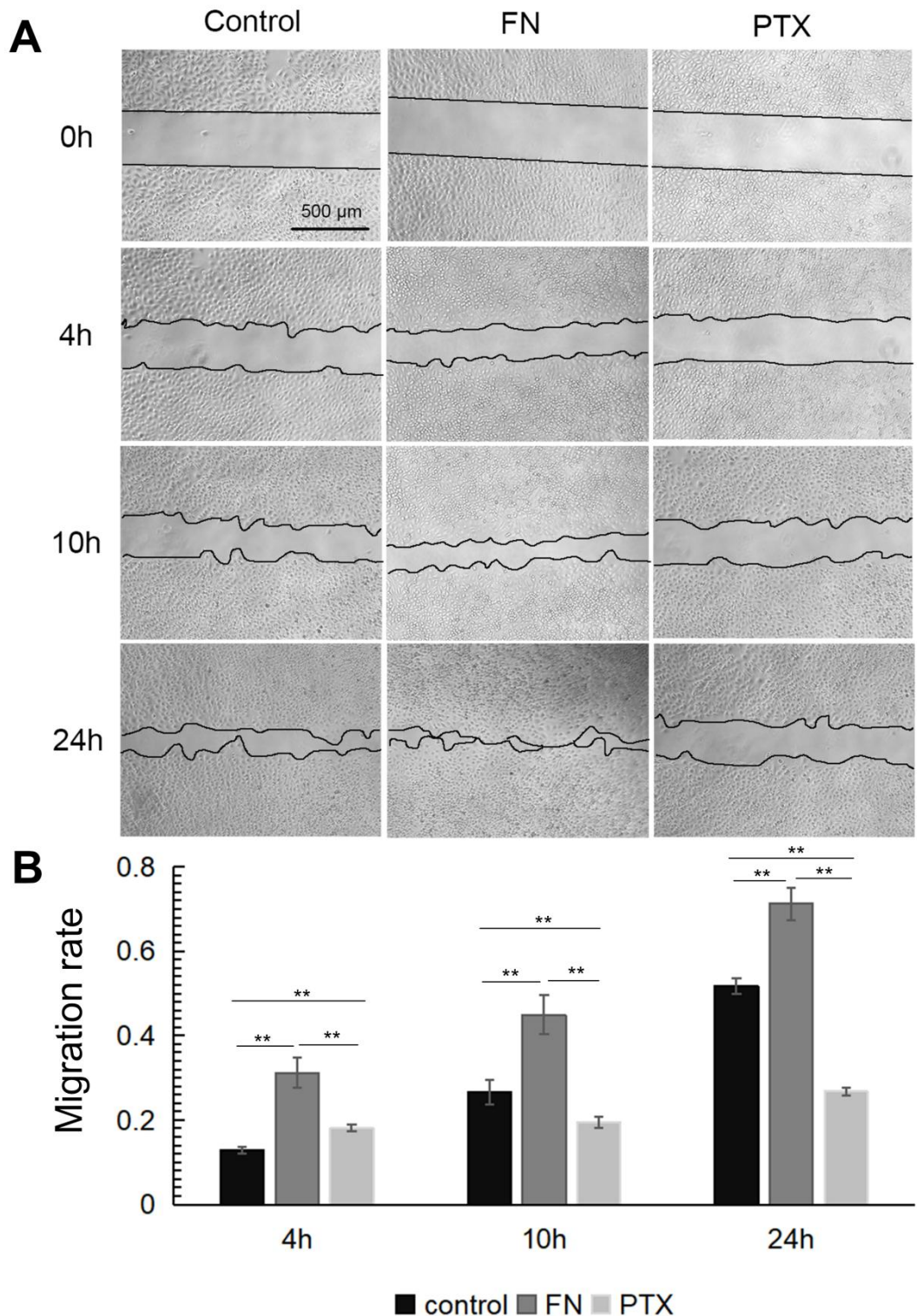

Figure 4. The effect of different coatings on HUVEC migration (scratch method). (A) Cell migration fluorescence microscopy images at different time points; cell magnification, 40×; scale bar, 500 μm; the black line indicates cell migration. (B) Quantitative analysis of the effects of the different coatings on the extent of HUVEC migration. **:p < 0.01.

### 3.5 VSMC proliferation and inhibition of the stents in vitro

To investigate VSMC growth on stent samples, the VSMC cytoskeleton and nucleus were traced. The cells in each group were analyzed via fluorescence microscopy after 3 days of incubation. (Figure 5A) In the bare scaffold group, smooth muscle cells grew throughout the scaffold. In the ADES group, smooth muscle cells grew better on the inner wall of the scaffold, while on the outer wall of the scaffold, smooth muscle cells grew in smaller numbers. In the UDES group, the number of cells on both the inner and outer walls of the scaffolds was lower, and the morphology was less full. The results demonstrated that the ADES group was superior to the UDES group in terms of inhibiting smooth muscle cell proliferation on the outer wall while promoting cell proliferation on the inner wall. (Figure 5B)

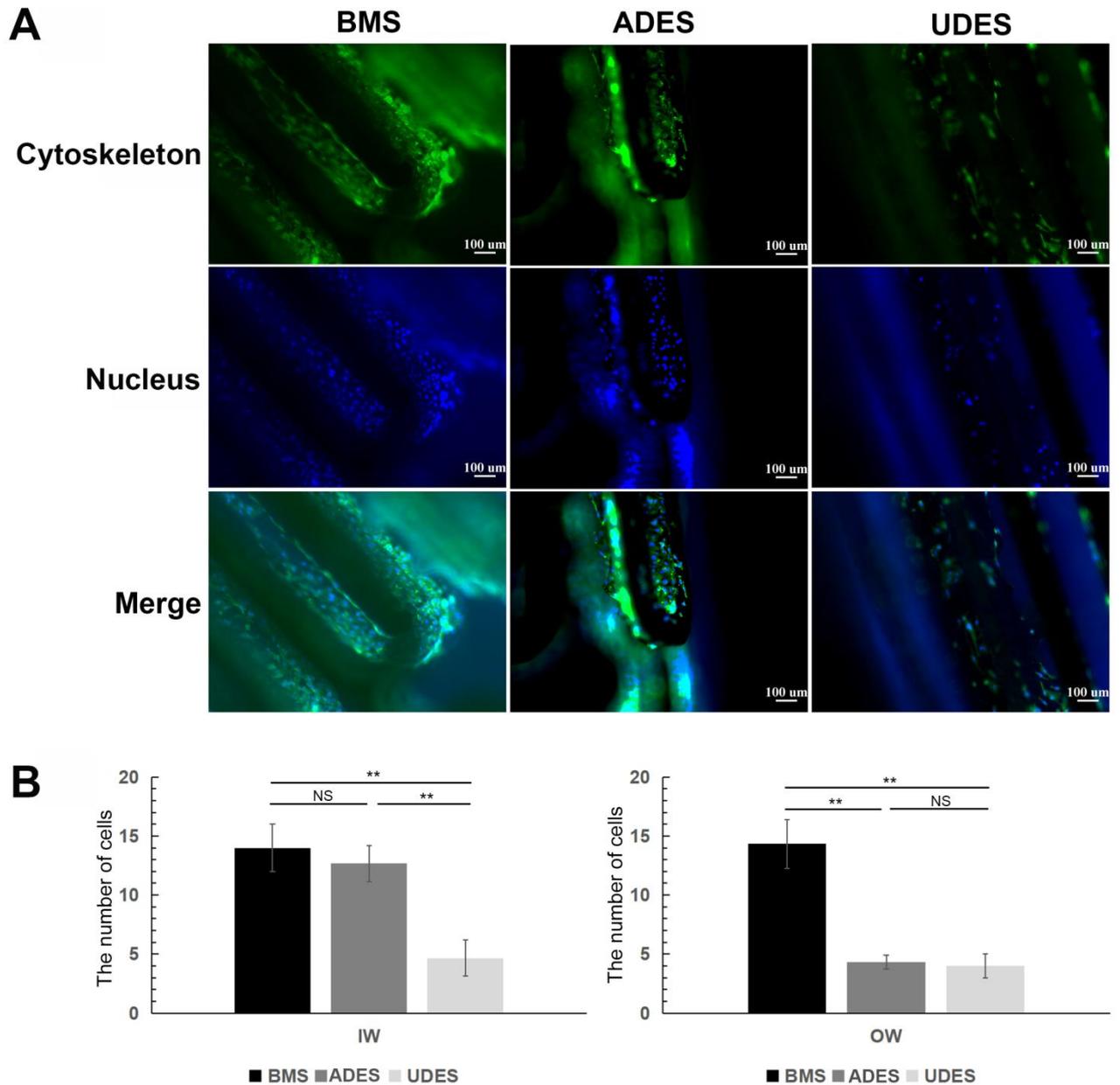

Figure 5. (A) Cytoskeleton immunofluorescence-labeled images showing VSMC growth on the stent wall. Green fluorescence indicates the cytoskeleton; blue fluorescence indicates the nucleus. Bar: 100 μm. (B) Number of cells counted per field of view. NS: not significant; **: $p < 0.01$.

### 3.6 Kinetic drug release characteristics in vitro

The contents of PTX and FN in the ADES films were 50.21±0.8241 μg and 40.00±1.0286, respectively. The contents of PTX and FN in the UDES films were 97.94±2.4344 μg and 39.74±1.1629 ng, respectively. The data were calculated from the mean total drug loading of the four stents. ADES and UDES were placed in a hemodynamic simulation system at a flow rate of 15 dyn/cm$^2$. (Figure 6A) The kinetic drug release data showed that the drug release patterns in the inner walls of the ADES and UDES were extremely different, while the drug release patterns in the outer wall were basically similar. The inner wall of the ADES mainly released FN, especially in the first 24 hours, and almost no PTX release was detected. During the first 7 days, only low amounts of PTX were also detected, which should have permeated from the outer wall. In contrast, the inner wall of the stent mainly released PTX from the UDES, and in the first 24 hours, almost no release of the FN was detected. The release of FN was consistently small and slow because it was covered by an antiproliferative layer. (Figure 6B) These results support the successful design of ADES for targeted drug release. Such a tissue-targeted pattern is important for early endothelial regeneration after stent implantation. As a result, this drug release pattern makes ADES superior to conventional UDES in inhibiting restenosis and promoting reendothelialization.

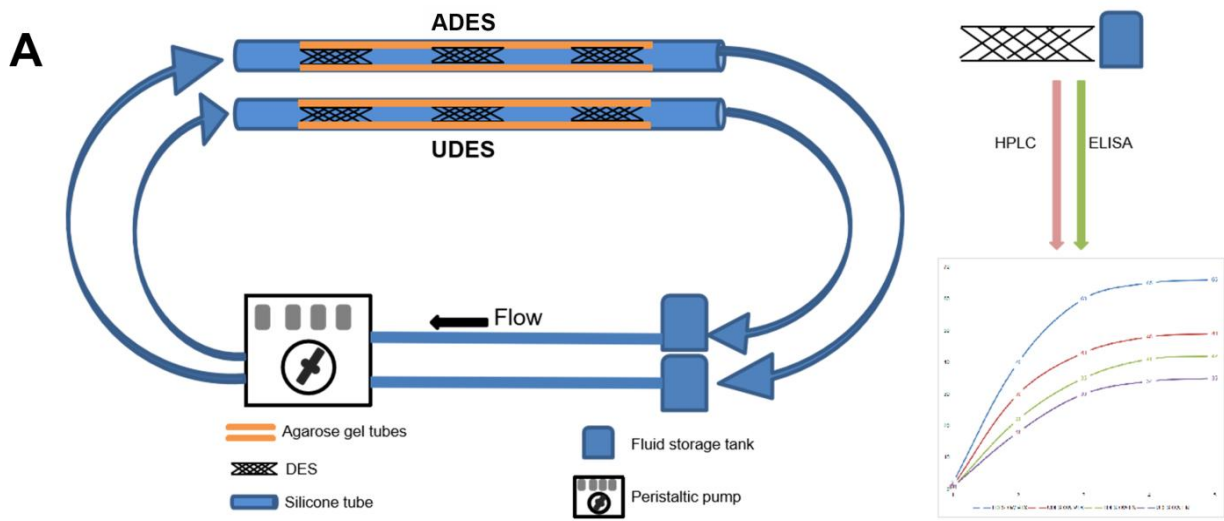

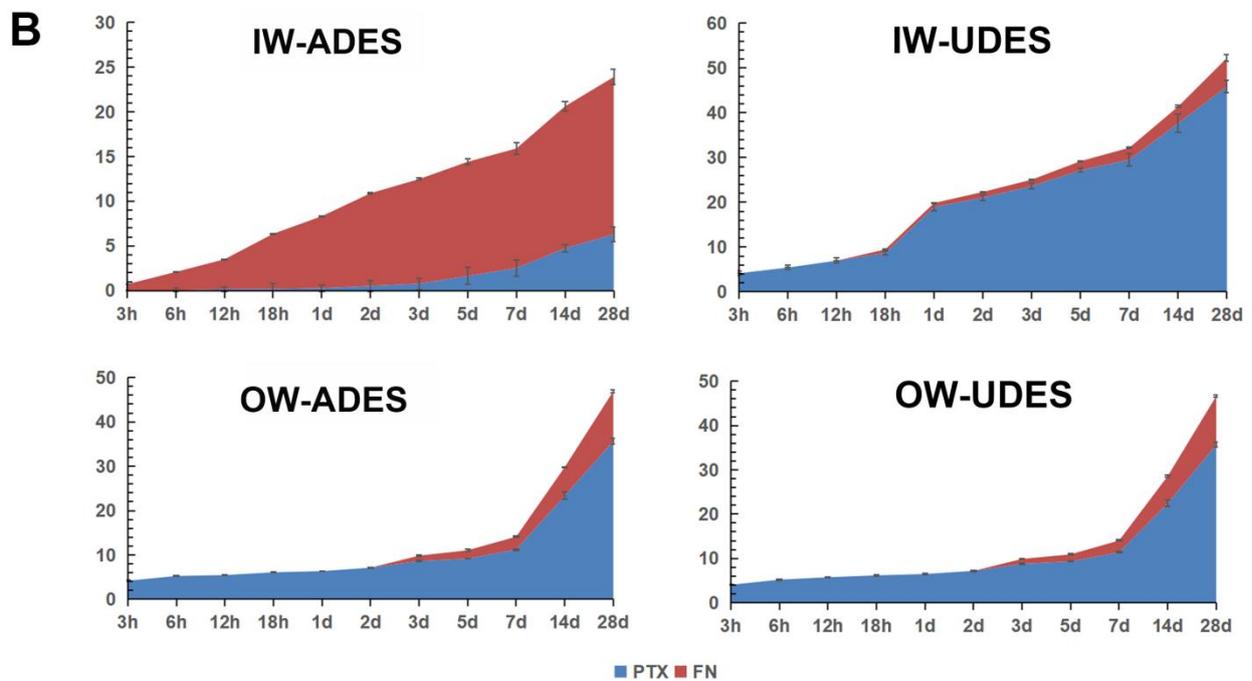

Figure 6. (A) The hemodynamic simulation system. Stent samples were placed in agarose gel tubes, luminal fluid was collected for stent inner wall drug release assays, and agarose gel tubes were collected for stent outer wall drug release assays. (B) The drug release patterns of the inner and outer walls of ADES and UDES.

**3.7 Stent re-endothelialization**

One week after stent implantation, the inner surface of the stent in both the UDES and BMS groups was exposed to the vessel intima with no endothelial cell growth, while partial endothelial repair was observed on the inner surface of the stent in the ADES group. Two weeks after stent implantation, partial endothelial repair was observed in the BMS group, most of the endothelialization was achieved in the ADES group, and only a small amount of endothelialization was observed in the UDES group. Four weeks after stent implantation, complete endothelialization was achieved in the ADES group, most of the endothelialization was achieved in the BMS group, and some of the endothelialization was achieved in the UDES group. Twelve weeks after stent implantation, the stent filaments in the BMS group were completely covered by the vessel intima and could not be observed; the stents in the ADES group were well endothelialized with a smooth intima; and the stents in the UDES group were endothelialized. (Figure 7)

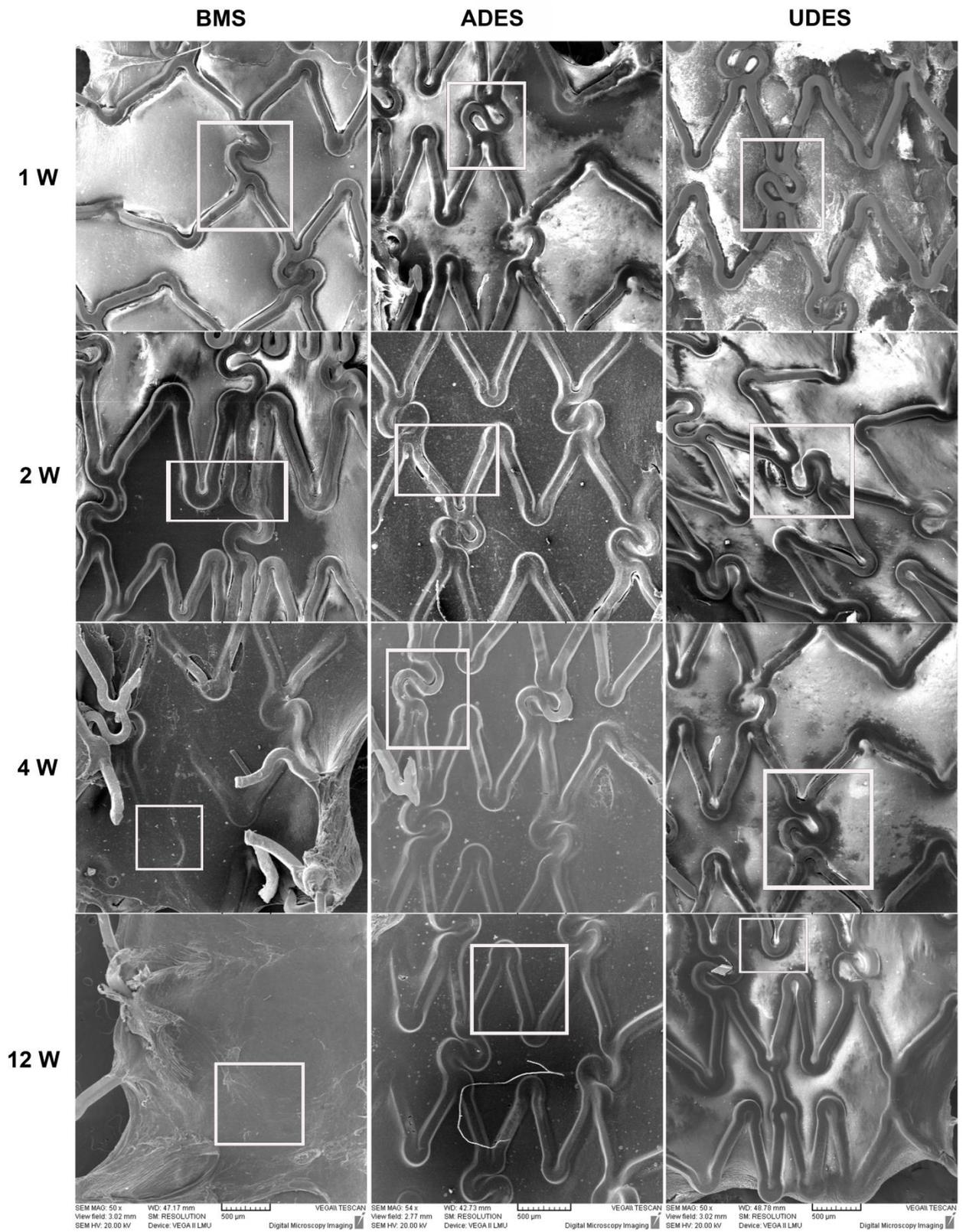

Figure 7. SEM images of stent endothelialization morphology.

**4 Discussion**

The development and evaluation of ADES represent a significant step forward in the quest for more effective and safer PCI. Our findings suggest that by tailoring drug delivery to the specific biological environments encountered by different surfaces of the stent, it is possible to enhance therapeutic outcomes while minimizing adverse effects.

One of the key findings from our in vitro tests was the respectable level of hemocompatibility exhibited by the ADES. This is crucial, as any foreign material introduced into the bloodstream must minimize interactions with blood components to avoid triggering thrombotic events.[15] The hemocompatibility of the ADES suggests that its design does not elicit an undue immune response or promote clot formation, which is a critical requirement for any intravascular device.

The drug release profiles observed in our study further validate the concept of asymmetric drug delivery. By demonstrating targetability in both the wall and lumen of artificial blood arteries, the ADES shows promise in delivering drugs to the precise locations where they are needed most. This targeted approach could potentially lead to improved efficacy in preventing restenosis and promoting endothelialization without causing systemic side effects associated with non-targeted drug release.

The significant endothelialization observed on the surface of stented segments implanted with ADES in the rabbit carotid artery model is particularly noteworthy. Endothelial cells play a pivotal role in maintaining vascular homeostasis, and their rapid proliferation over the stent surface is essential for preventing late stent thrombosis and ensuring long-term patency.[16] The fact that the ADES facilitated this process more effectively than traditional uniform-coating DES supports our hypothesis that asymmetric drug delivery can enhance healing responses post-implantation.

It is important to consider these results within the context of existing literature. While previous studies have explored various modifications to DES designs, including biodegradable polymers and different drug combinations, [17, 18] few have focused specifically on creating an asymmetry between drug types delivered to the inner versus outer surfaces of the stent. Our study contributes novel insights into how such asymmetric designs can be realized through advanced manufacturing techniques like ultrasonic atomizing spraying.

Furthermore, the ultrasonic atomizing spraying technique employed in the fabrication of ADES offers a versatile platform that accommodates a wide range of stent materials without imposing stringent requirements. This method is adaptable to various metal alloys, polymers, sizes, and shapes of stents, thereby enhancing its applicability across different cardiovascular intervention scenarios. Moreover, the technique exhibits remarkable flexibility regarding the substances being sprayed; any material capable of dissolving in a suitable solvent can be effectively applied by adjusting the atomization parameters, enabling precise control over drug deposition and release kinetics.

However, it is also essential to acknowledge the limitations of our study. Animal models, although informative, do not fully recapitulate the complexities of human cardiovascular disease and healing processes.[19, 20] Therefore, caution must be exercised when extrapolating these findings to clinical scenarios. Future research should include larger animal studies and eventually move toward clinical trials to confirm the safety and efficacy of ADES in humans.

**5 Conclusion**

In conclusion, our study presents compelling evidence that ADES holds considerable promise as a new generation of cardiovascular intervention devices. By leveraging asymmetric drug delivery strategies, these stents have the potential to address some of the unmet needs in current PCI treatments, ultimately leading to better patient outcomes. Further investigations are warranted to optimize stent design, refine drug selection, and establish the long-term benefits and risks associated with this innovative technology.

## Acknowledgments

This work was supported by grants from the National Key R & D Project of China (2023YFB3810100) and the Science and Technology Innovation Project of Jinfeng Laboratory, Chongqing, China (jfkyjf202203001), the Fundamental Research Funds for the Central Universities (2024CDJCGJ-016, 2023CDJYGRH-ZD03), and China Postdoctoral Science Foundation (2023MD734198). The authors also thank Professor Chaojun Tang of Soochow University for her valuable suggestions on the revision of this article. The authors are also thankful to the First Batch of Key Disciplines on Public Health in Chongqing and the Public Experiment Center of the State Bioindustrial Base (Chongqing), China.

## Competing interests

The authors declare no competing interests.